\documentclass[letterpaper,journal]{IEEEtran}

\usepackage{algorithmic}
\usepackage[T1]{fontenc}
\usepackage[utf8]{inputenc}
\usepackage{algorithm}
\usepackage{array}
\usepackage[caption=false,font=normalsize,labelfont=sf,textfont=sf]{subfig}
\usepackage{textcomp}
\usepackage{stfloats}
\usepackage{hyperref}
\usepackage{cite}
\usepackage{siunitx}

\usepackage{booktabs}

\usepackage{url}
\usepackage{verbatim}
\usepackage{graphicx}

\begin{document}

\title{First positronium imaging using $^{44}$Sc with the J-PET scanner: a case study on the NEMA-Image Quality phantom}

\author{Manish Das, Sushil Sharma, Ermias~Yitayew~Beyene, Aleksander Bilewicz, Jarosław Choiński, Neha Chug, Catalina Curceanu, Eryk~Czerwiński, Kavya~Valsan~Eliyan, Jakub Hajduga, Sharareh Jalali, Krzysztof Kacprzak, Tevfik Kaplanoglu, Łukasz Kapłon, Kamila~Kasperska, Aleksander Khreptak, Grzegorz Korcyl, Tomasz Kozik, Karol Kubat, Deepak~Kumar, Anoop~Kunimmal Venadan, Edward~Lisowski, Filip Lisowski, Justyna~Medrala-Sowa, Simbarashe Moyo, Wiktor~Mryka, Szymon Nied{\'z}wiecki, Piyush Pandey, Szymon Parzych, Alessio Porcelli, Bartłomiej Rachwał, Elena Perez del Rio, Martin Rädler, Axel Rominger, Kuangyu Shi, Magdalena Skurzok, Anna~Stolarz, Tomasz~Szumlak, Pooja Tanty, Keyvan~Tayefi~Ardebili, Satyam Tiwari, Rafał~Walczak, Ewa Ł. Stępień and Paweł Moskal


\thanks{Manuscript received xxxx; revised xxxx; accepted xxxx. Date of publication xxxx; date of current version xxxx.
We acknowledge support from the National Science Centre of Poland through grants MAESTRO no. 2021/42/A/ST2/00423 (P.M.), OPUS no. 2021/43/B/ST2/02150 (P.M.), OPUS24+LAP no. 2022/47/I/NZ7/03112 (E.S.), SONATA no. 2023/50/E/ST2/00574 (S.S.) and SONATA no. 2020/38/E/ST2/00112 (E.P. del R.), the Ministry of Science and Higher Education through grant no. IAL/SP/596235/2023 (P.M.), the SciMat and qLife Priority Research Areas budget under the program Excellence Initiative – Research University at Jagiellonian University (P.M. and E.S.), the Research Support Module as part of the Excellence Initiative – Research University program at Jagiellonian University (M.D.). We also acknowledge Polish high-performance computing infrastructure PLGrid (HPC Center: ACK Cyfronet AGH) for providing computer facilities and support within computational grant no. PLG/2024/017688.\\(\textit{Corresponding authors:} Manish Das: manish.das@doctoral.uj.edu.pl, Paweł Moskal: p.moskal@uj.edu.pl)

This work did not involve human subjects or animals in its research. \\}

\thanks{Manish Das, Sushil Sharma, Ermias Yitayew Beyene, Neha Chug, Eryk~Czerwiński, Kavya~Valsan~Eliyan, Sharareh Jalali, Krzysztof Kacprzak, Tevfik Kaplanoglu, Łukasz Kapłon, Kamila Kasperska, Aleksander Khreptak, Grzegorz Korcyl, Tomasz Kozik, Karol Kubat, Deepak~Kumar, Anoop Kunimmal Venadan, Justyna Medrala-Sowa, Simbarashe Moyo, Wiktor Mryka, Szymon Nied{\'z}wiecki, Piyush Pandey, Szymon Parzych, Alessio Porcelli, Elena Perez del Rio, Martin Rädler, Magdalena Skurzok, Pooja Tanty, Keyvan~Tayefi~Ardebili, Satyam Tiwari, Ewa Ł. Stępień and Paweł Moskal are with the Marian Smoluchowski Institute of 
Physics and the Center for Theranostics, Jagiellonian University, 31-007 Kraków, Poland 

Aleksander Bilewicz and Rafał~Walczak are with with the Institute of Nuclear Chemistry and Technology, Warsaw, Poland.

Jarosław Choiński and Anna~Stolarz are with Heavy Ion Laboratory, University of Warsaw, Warsaw, Poland.

Catalina Curceanu is with INFN, Laboratori Nazionali di Frascati CP 13, Via E. Fermi 40, 00044, Frascati, Italy.

Jakub Hajduga,  Bartłomiej Rachwał, Tomasz Szumlak are with AGH University of Krakow, Poland.

Edward~Lisowski and Filip Lisowski are with Cracow University of Technology, 31-864 Kraków, Poland.

Axel Rominger and  Kuangyu Shi are with the Department of Nuclear Medicine, Inselspital, Bern University Hospital, University of Bern, 3010, Bern, Switzerland. }}

\markboth{IEEE TRANSACTIONS ON RADIATION AND PLASMA MEDICAL SCIENCES, VOL. X, NO. X, MAY XXXX}%
{Shell \MakeLowercase{\textit{Das et al.}}: A Sample Article Using IEEEtran.cls for IEEE Journals}


\maketitle

\begin{abstract}
Positronium Lifetime Imaging (PLI), an emerging extension of conventional positron emission tomography (PET) imaging, offers a novel window for probing the submolecular properties of biological tissues by imaging the mean lifetime of the positronium atom. Currently, the method is under rapid development in terms of reconstruction and detection systems. Recently, the first in vivo PLI of the human brain was performed using the J-PET scanner utilizing the $^{68}$Ga isotope. However, this isotope has limitations due to its comparatively low prompt gamma yields, which is crucial for positronium lifetime measurement. Among alternative radionuclides, $^{44}$Sc stands out as a promising isotope for PLI, characterized by a clinically suitable half-life (4.04 hours) emitting 1157 keV prompt gamma in 100\% cases after the emission of the positron. This study reports the first experimental demonstration of PLI with $^{44}$Sc, carried out on a NEMA-Image Quality (IQ) phantom using the Modular J-PET tomograph—the first plastic scintillators-based PET scanner. 
\end{abstract}


\section{Introduction}
\IEEEPARstart{P}{ositron} Emission Tomography (PET) is an advanced nuclear imaging modality enabling the quantitative assessment of physiological and biochemical processes in vivo by detecting pairs of coincident gamma photons 
produced during positron-electron annihilation events~\cite{Phelps661,Alavi:2021}. Recently, the method of positronium imaging was invented \cite{IEEEpositronium-imaging, Moskal2025IEEE} that enables imaging of positronium properties during the PET diagnosis. In the first clinical studies, the images of the mean positronium lifetime were demonstrated using the multi-photon J-PET scanner~\cite{doi:10.1126/sciadv.adp2840}. Next, positronium lifetime measurements were performed using Biograph Vision Quadra for reference materials~\cite{Steinberger2024} and humans~\cite{Mercolli2024.10.19.24315509}.

Positronium Lifetime Imaging (PLI), relies on the fact that, in the human body, in almost 40$\%$ of cases, a metastable bound state (positronium atom, Ps) is formed. It consists of a positron emitted from an administered radiopharmaceutical and an electron from its molecular enviroment. Ps can be formed in one of the two spin configurations: para-positronium (pPs) or ortho-positronium (oPs), and according to spin-statistics, it is produced in a 1:3 ratio~\cite{EJNMMI2020}. In the vacuum, the mean lifetime of oPs is equal to 142~ns, however, it is shortened to a few nanoseconds (1.4 to 2.9~ns \cite{Ahn-Gidley-colagen-2021, doi:10.1126/sciadv.abh4394, Jean2006, Chen2012, Jasinska2017, Zgardzinska2020, Karimi2023, EJNMMI2023, Moyo2022Feasibility}) within the environment of biological tissues, reflecting its sensitivity to the local molecular environment~\cite{RMP2023}. This observation advocates that imaging the Ps lifetime could serve as a novel biomarker for probing the submolecular characteristics and, potentially, for disease characterization at the submolecular level~\cite{EJNMMI2023}. Therefore, PLI stands as a new method that may provide additional information in comparison to conventional PET/CT scans~\cite{Moskal2025IEEE}.

The oPs produced within tissue can be influenced due to the onset of several mechanisms leading to the shortening of its lifetime. One of such mechanisms is known as the pick-off process, in which the Ps picks-off an e$^-$ from the surroundings and annihilates, mainly into two-photons~\cite{jean:2003principles}. Another dominant mechanism is due to the presence of oxygen molecules (O$_2$) in tissues. The lifetime of oPs is reduced either by spin-exchange with paramagnetic (O$_2$), flipping oPs into pPs which rapidly annihilates or through the oxidation under reactive oxidizing environments (e.g., presence of free radicals) where oPs undergoes an electron transfer reaction~\cite{RMP2023, Gidley2006}.

Several research groups have conducted feasibility studies to explore the clinical use of PLI as reported in~\cite{PMB2019, doi:10.1126/sciadv.abh4394, doi:10.1126/sciadv.adp2840, Qi-positronium-2022, Qi-positronium-2023, Huang2025Fast, Huang2025Statistical,Steinberger2024,ManishBAMS,Shibuya2020,  Moskal:2022A, Shibuya:2022, Mercolli2024.10.19.24315509,Takyu_2024,TAKYU2024169514, Huang_mic, Huangjnumed.125.270130}. A PET scan detects 511 keV photon pairs that stem from both direct positron annihilations and Ps formation annihilations, but the latter have not been fully explored. In a conventional scanner, the detection time of annihilation photons gives an estimate of Ps decay time but  determining its lifetime requires information about when Ps was formed. A positron-emitting ($\beta^+$) source with an immediate  prompt gamma release after $\beta^+$ decay can meet this requirement.

In physics experiments, $^{22}$Na is the isotope of choice, e.g. for fundamental symmetries in the decays of Ps atoms~\cite{Moskal:2021kxe,Moskal2025Nonmaximal, Moskal2024Discrete}. The $^{22}$Na emits a 1275~keV prompt gamma ray in almost every $\beta^+$ decay (99.94$\%$ branching ratio), thus providing a one-to-one coincidence between positron emission and prompt gamma release, where the prompt gamma marks the formation time of Ps. This intrinsic timing marker makes $^{22}$Na well suited for in-vitro and ex-vivo positronium lifetime studies in the laboratory~\cite{Ahn-Gidley-colagen-2021, doi:10.1126/sciadv.abh4394, Jean2006, Chen2012, Jasinska2017, Zgardzinska2020, Karimi2023, EJNMMI2023, Moyo2022Feasibility}. The J-PET collaboration used this capability to conduct an ex vivo investigation of human cardiac myxoma and adipose  tissues by placing $^{22}$Na source between each of these samples and measuring the oPs mean lifetimes. The results showed significant differences between myxoma at 1.9~ns and fat at  2.6~ns thus proving PLI's ability to detect variations in biological tissues~\cite{EJNMMI2023, doi:10.1126/sciadv.abh4394}. However, $^{22}$Na cannot be used for imaging of humans due to its long half-lifetime of 2.60 years and uptake in bones~\cite{Triffitt1970boneuptake}.

In general, multiple obstacles emerge when moving from controlled ex-vivo measurements to in-vivo studies because physiological motion, tissue heterogeneity, and background activity make precise lifetime determination challenging. The implementation of in vivo PLI requires a PET system with subnanosecond time-of-flight resolution, and a radionuclide with high branching ratio for $\beta^+$ decay followed by the emission of prompt gamma. In addition, the radionuclide must have a suitable half-life for clinical use and must be manufactured through standard protocols (see  Refs.~\cite{ManishBAMS,Sitarz2020}). 

In the first clinical trial, J-PET performed human brain PLI using $^{68}$Ga-PSMA, demonstrating the clinical potential of PLI through successful in-vivo mean lifetime imaging of oPs~\cite{doi:10.1126/sciadv.adp2840}. Despite these advances with $^{68}$Ga radionuclide, the fact that in only 1.34\% of cases the prompt gamma ray accompanies the 511~keV photons imposes a major limitation for PLI.

\begin{figure}[ht]
    \centering
        \includegraphics*[width=0.5\textwidth]{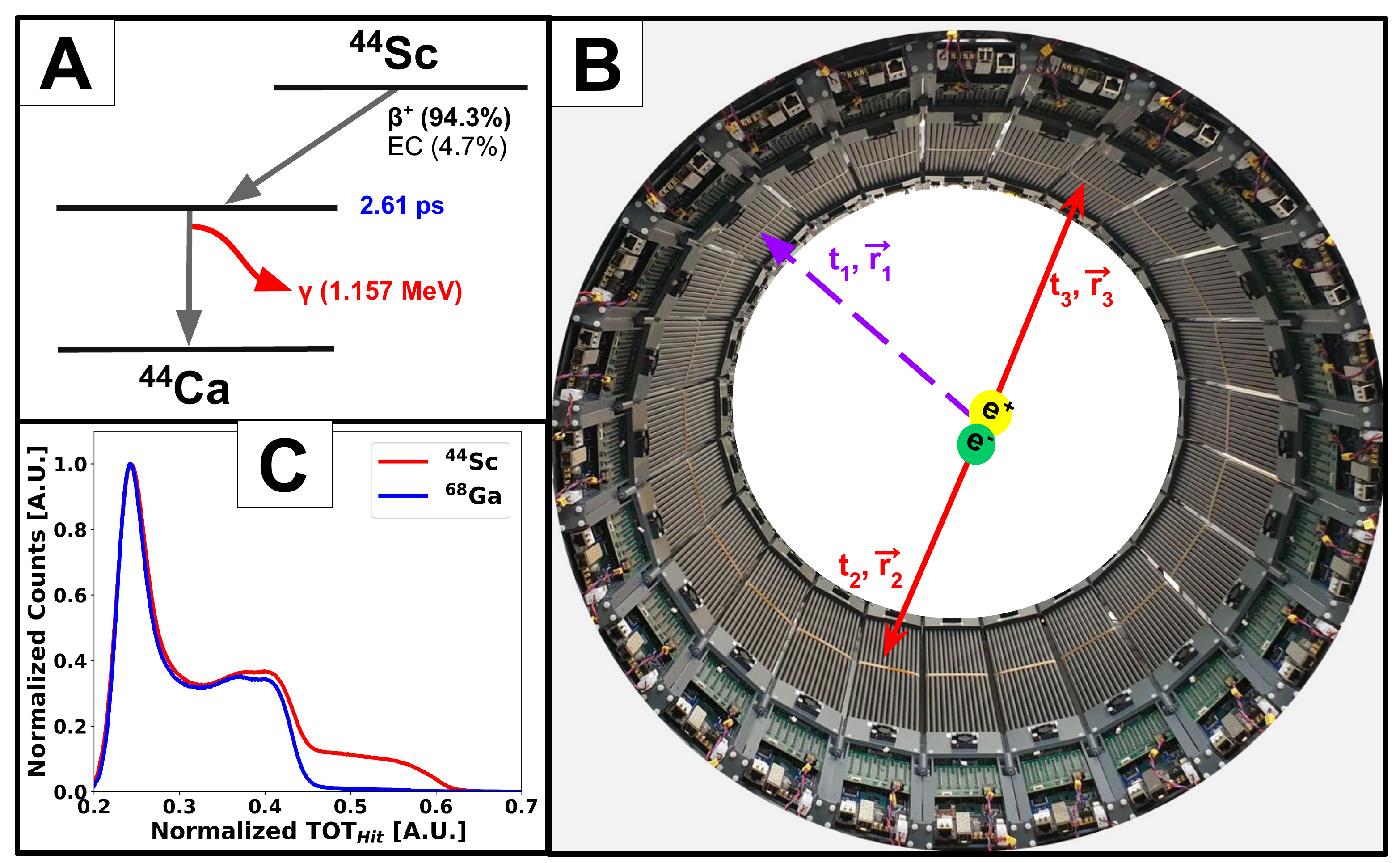}
    \caption{\textbf{(A)} The decay schemes for the $^{44}$Sc isotope. $\beta^{+}$ denotes the positron yield, EC indicates electron capture contributions, and $\gamma$ represents the prompt gamma with energy indicated in the paranthesis. Additionally, the delay time is presented in blue text for clarity. The delay time denotes the average time between a positron's emission and a prompt gamma's emission. \textbf{(B)} The event definition of one prompt gamma ($t_1,\vec{r}_1$) and two annihilation photons (($t_2,\vec{r}_2$) and ($t_3,\vec{r}_3$)) is depicted in the modular J-PET, requisite for positronium lifetime estimation. \textbf{(C)} Experimental Time-Over-Threshold (TOT) spectra for $^{68}$Ga and $^{44}$Sc, showing an increased prompt gamma yield for $^{44}$Sc in the modular J-PET scanner. 
\label{introduction}
    }
\end{figure}

In contrast, $^{44}$Sc emerges as the best potential candidate for PLI \cite{pet-clinics} with a favorable decay profile: a clinically compatible half-life of 4.04~h~\cite{Duran2022Half}, an ultrashort de-excitation delay of 2.61~ps, and 94.3\% of decays resulting in the emission of a positron followed by the high-energy (1157 keV) prompt gamma~\cite{ManishBAMS}. This corresponds to a prompt gamma yield accompanied with the positron emission roughly 75 times greater than that of $^{68}$Ga~\cite{ManishBAMS}. The decay scheme of $^{44}$Sc is shown in Fig.~\ref{introduction}(A).

Although the medical application of \textsuperscript{44}Sc has not yet been established in clinical practice, its potential has been studied in detail in both the preclinical and clinical settings~\cite{Domnanich2016, Koumarianou2012, Muller2013, Eigner2012, Hernandez2014, Chakravarty2014, Szucs2022, Majkowska2011Macrocyclic, Pruszynski2012Radiolabeling, Duran2022Half}. The performance of PET imaging using $^{44}$Sc has been shown through phantom studies~\cite{Sc5_Lima2020First}, comparative evaluations of radionuclides \cite{Sc1_lima}, and preclinical radiotheranostic applications~\cite{Sc2_Umbricht201744Sc,Sc4_Muller2014Promising}. In addition, first-in-human trials have shown its  applicability for imaging neuroendocrine tumors and prostate cancer~\cite{Sc6_Zhang2019From,Sc7_Eppard2017Clinical,Sc9_first_Singh267, Khawar201844Sc} and production methods suitable for clinical translation have been optimized~\cite{Sc3_van2020Developments}.

Presently, only six PET systems, the J-PET scanner in Cracow, Poland~\cite{doi:10.1126/sciadv.adp2840}, the Prism-PET scanner in New York City, USA~\cite{Huang2025Fast}, the Biograph Vision Quadra in Bern, Switzerland~\cite{Steinberger2024, Mercolli2024.10.19.24315509, Prenosil-quadra-2022}, the PennPET Explorer in  Philadelphia, USA  \cite{Dai-PennPET-NEMA, Huang_mic, Karp136, Huangjnumed.125.270130}, the NeuroEXPLORER (NX) brain PET scanner in USA \cite{Samanta2023Feasibility} and the  brain-dedicated TOF-PET scanner VRAIN in Chiba, Japan  \cite{Takyu_2024,  TAKYU2024169514}-are capable of measuring the lifetime of positronium. In recent years, several reconstruction algorithms for PLI have also been proposed  \cite{Huang2025Fast, Huang2024High , Huang2025Statistical, Berens2024analytic, Qi-positronium-2022, Qi-positronium-2023, Chen2023, Shopa-Kamil-Bams2023, Huang_mic, Huangjnumed.125.270130, Chen2024, Moskal2025IEEE}. 

Recent investigations using the Biograph Vision Quadra \cite{Mercolli2024.10.19.24315509, Steinberger2024} have revealed energy-detection limits for high-energy prompt gammas. The system records photons below 950~keV but cannot resolve energy depositions above 726~keV. This results in many events being lost, and the imprecise identification of $\gamma$-rays can increase background counts. Although in the case of $^{124}$I, lower energy prompt gammas remain detectable, using $^{68}$Ga and $^{44}$Sc suffers substantial photon losses under these constraints. In contrast, the J-PET scanner, the Prism-PET scanner, the PennPET Explorer and the brain-dedicated VRAIN system used for PLI~\cite{Dai-PennPET-NEMA, Huang_mic, Huangjnumed.125.270130, Takyu_2024, TAKYU2024169514, doi:10.1126/sciadv.adp2840}, do not have this constraint of energy detection.

Fig.~\ref{introduction}(B) shows the modular J-PET scanner~\cite{FaranakBAMS, FaranakBAMS2024} which was used to demonstrate the first positronium images~\cite{doi:10.1126/sciadv.abh4394, doi:10.1126/sciadv.adp2840}. This scanner was used in the study presented here. Fig.~\ref{introduction}(C) presents the experimental spectra of Time-Over-Threshold (TOT) values (which is a measure of the energy deposition~\cite{Sharma:2020}). One can see that for TOT values above 0.45, corresponding to the registration of prompt gamma, the signal for $^{44}$Sc is higher than that for $^{68}$Ga. 

In this work, we present the first experimental demonstration of PLI with $^{44}$Sc carried out on a NEMA Image Quality phantom using the modular J-PET scanner~\cite{pet-clinics,doi:10.1126/sciadv.adp2840}, which supports multiphoton detection without energy restrictions on deexcitation photons (Fig. \ref{introduction}C).

\section{Methods}\label{sec2}

\subsection{Isotope preparation}\label{subsec0}

In this study, two radionuclides emitting positrons were investigated: Fluorine-18 ($^{18}$F) and Scandium-44 ($^{44}$Sc). 
These isotopes exhibit different characteristics: $^{18}$F decays predominantly by $\beta^{+}$ decay (97\%) and, to a lesser extend, by electron capture (3\%), without significant prompt gamma emission. In contrast, $^{44}$Sc undergoes $\beta^{+}$ decay in approximately 94.3\% and subsequently emits a 1157 keV prompt $\gamma$-ray with a yield of around 100\%~\cite{ManishBAMS}. The inherited differences in the prompt gamma emission serve to demonstrate the effectiveness of the selection criteria developed to identify events having additional prompt gammas for PLI. Moreover, 
this approach can be extended for effective implementation of simultaneous imaging of double isotopes as described in~\cite{Beyene2023Exploration}, although this is beyond the scope of the presented work.

The $^{18}$F was purchased from the commercial supplier VOXEL and $^{44}$Sc was produced by the Heavy Ion Laboratory of the University of Warsaw~\cite{HILTarget1, HILTarget2, ChoinskiBAMS}  via the $^{44}$Ca(p,n)$^{44}$Sc reaction. A natural calcium carbonate target ($\text{CaCO}_3$ tablets) was irradiated for 2 hours and 20 minutes using a 16.5 MeV proton beam with a current of 15 $\mu$A. Each tablet weighed 70 mg and was pressed onto
a black graphite pad, with a total weight of 342 mg per disc. After irradiation, the target disc was dissolved in diluted hydrochloric acid (HCl) to obtain a scandium solution. The solution was then filtered to remove graphite dust, separated, concentrated, and subsequently buffered to raise the pH from 0.5 to approximately 4.0.

\subsection{Phantom Study }\label{subsec1}

The NEMA IQ Phantom (Pro-Project Pro-NM NEMA NU2) was used in the study, which comprises six spherical inserts of 10, 13, 17, 22, 28 and 37 mm diameter and a lung-equivalent insert filled with styrofoam balls. The background compartment of the phantom was filled with demineralized water and contained no radioactivity.
\begin{figure}[htbp]
    \centering
        \includegraphics*[width=0.5\textwidth]{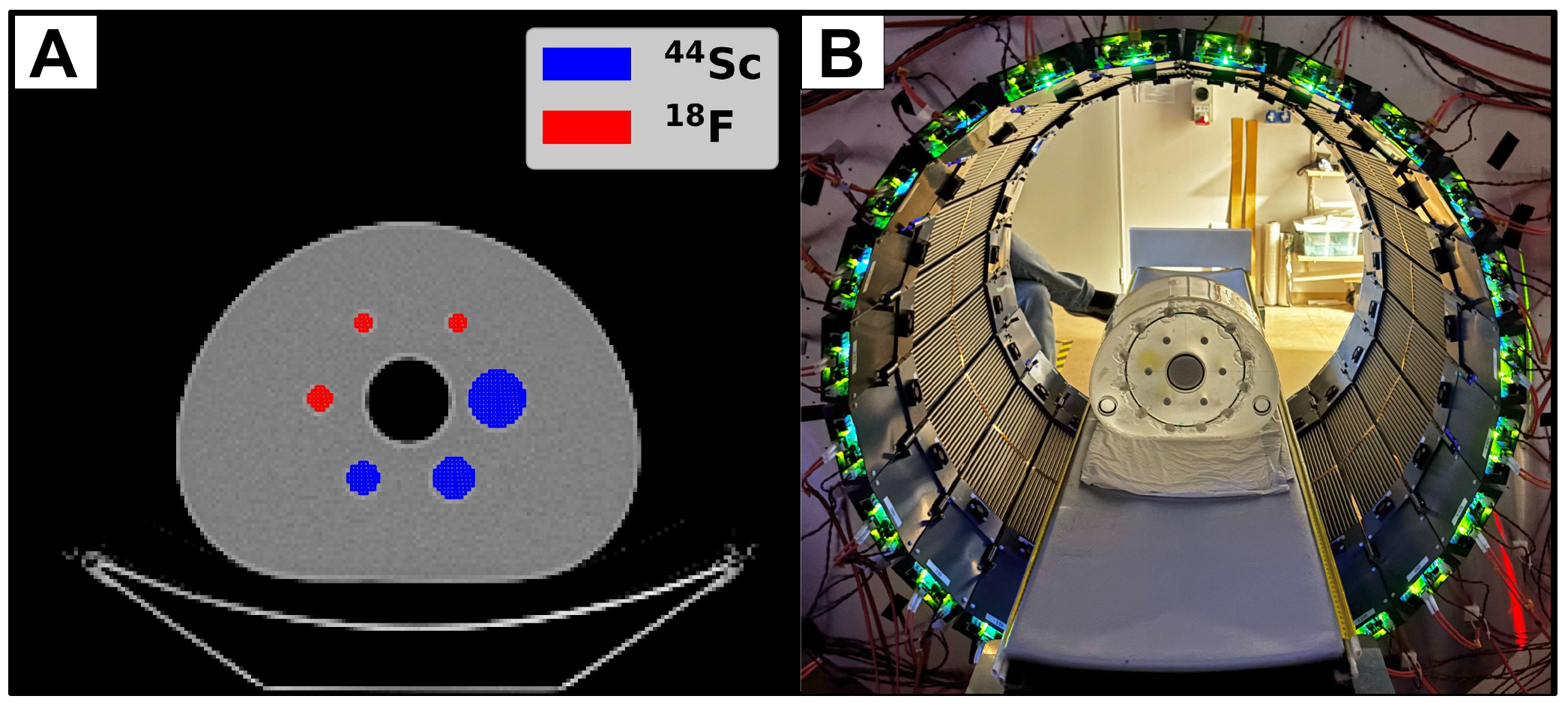}
    \caption{ \textbf{(A)} Transaxial CT scan of the NEMA IQ phantom schematically depicting the radiotracer distribution in the spheres, with those highlighted in red containing $^{18}$F and those in blue filled with $^{44}$Sc. \textbf{(B)} The NEMA IQ phantom positioned inside the modular J-PET detector. 
\label{method}
    }
\end{figure}

In the NEMA IQ phantom, shown in Fig. \ref{method}(A), spheres with diameters of 10 mm, 13 mm, and 17 mm were filled with $^{18}$F, while spheres with diameters of 22 mm, 28 mm, and 37 mm were filled with $^{44}$Sc, both mixed with water, at activity concentrations of 0.574 MBq/mL and 0.185 MBq/mL, respectively, measured at the start of the experiment. The phantom was then positioned inside the detector as shown in Fig. \ref{method}(B), and the data was acquired for a duration of 178 minutes.

\subsection{Event Selection for positronium imaging }\label{subsec2}

The event selection criteria, established in our previous work ~\cite{doi:10.1126/sciadv.abh4394,doi:10.1126/sciadv.adp2840}, provide a clear distinction between positronium-related signals and standard two-photon annihilation events used for standard PET imaging~\cite{doi:10.1126/sciadv.abh4394,doi:10.1126/sciadv.adp2840}. The principle of standard PET relies on the detection in coincidence of two back-to-back 511~keV annihilation photons ($\gamma_a$), while PLI requires both annihilation photons and a third additional prompt gamma ($\gamma_p$) used as start signal for Ps formation. The J-PET scanner operates in trigger-less mode allowing multiphoton detection simultaneously. The data acquisition and signal processing methods used in J-PET are explained in \cite{doi:10.1126/sciadv.adp2840,  Niedzwiecki:2017nka, korcyl_ieee}. 

\begin{figure}[htbp]
    \centering
        \includegraphics*[width=0.5\textwidth]{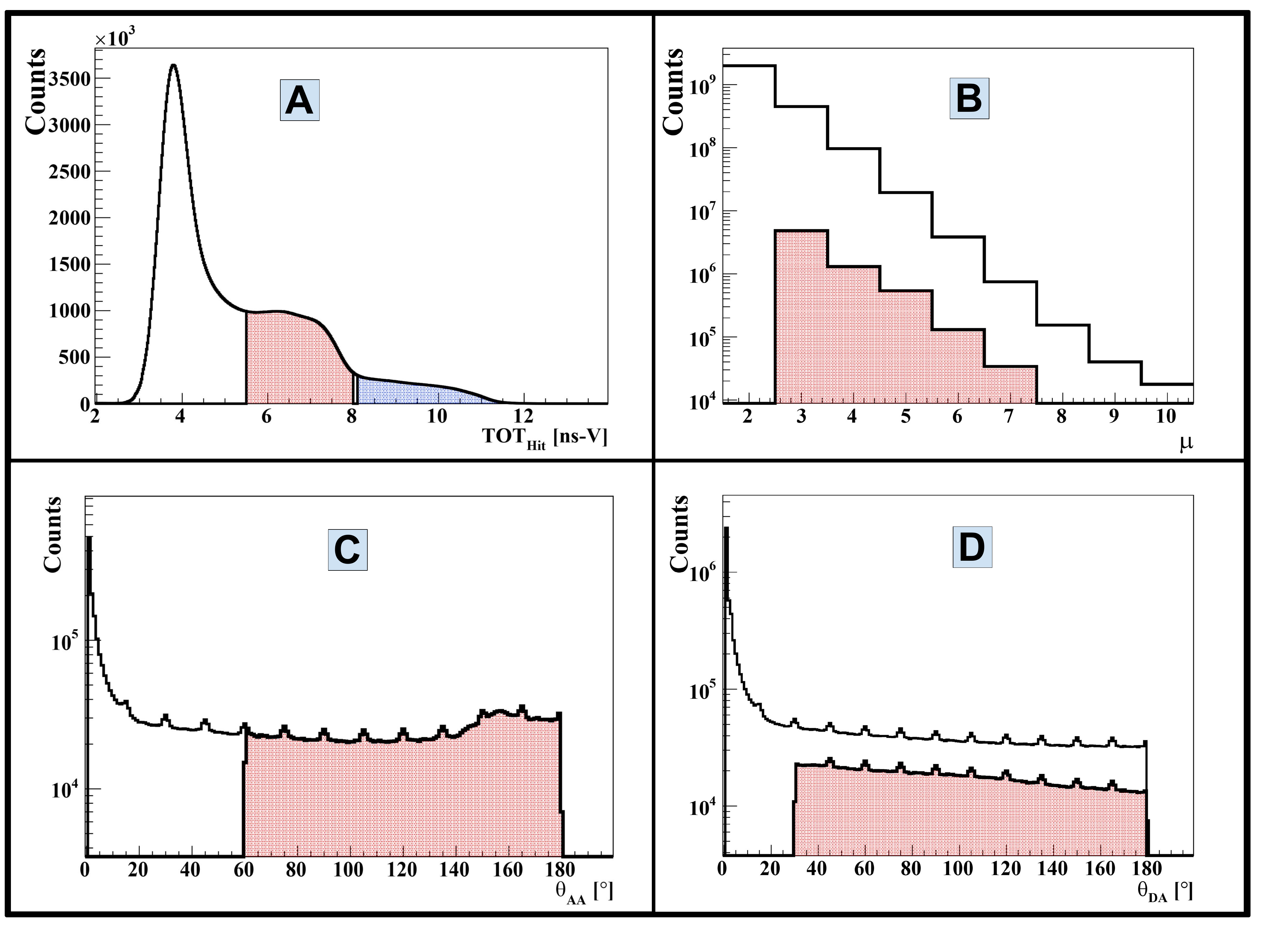}
    \caption{{\bf{(A)}} Distribution of time-over-threshold (TOT$_{Hit}$) for photon identification, with annihilation photons (red) and prompt gammas (blue) marked by distinct ranges. {\bf{(B)}} Hit multiplicity ($\mu$) distribution for events, with the red-shaded histogram highlighting selected events containing exactly two annihilation photons and one prompt gamma. For $\mu = 3$, selected three hits correspond to two photons in the annihilation region and one in the prompt region of the TOT$_{Hit}$ distribution. For $\mu > 3$, events are selected with exactly two hits in the annihilation region and one hit in the prompt region, while additional hits in $\mathrm{TOT}_{\mathrm{hit}} < 5.5 \mathrm{ns\cdot V}$ or $\mathrm{TOT}_{\mathrm{hit}}\in(8-8.1) \mathrm{ns\cdot V}$ are discarded. {\bf{(C)}} Distribution of the relative angle ($\theta_{AA}$) between annihilation photon vectors $\vec{r}_2$ and $\vec{r}_3$ (per Fig.~\ref{introduction}B), with $\theta_{AA} \geq 60^\circ$ (red) as the selection criterion. {\bf{(D)}} Distribution of the relative angle ($\theta_{DA}$) between prompt gamma vector $\vec{r}_1$ and annihilation photon vectors $\vec{r}_2$, $\vec{r}_3$ (per Fig.~\ref{introduction}B), with $\theta_{DA} \geq 30^\circ$ (red) as the restriction.
    \label{selection}
    }
\end{figure}

In this work, we classified events into two categories based on the radionuclide type: $^{18}$F produces two back-to-back 511 keV photons, and $^{44}$Sc emits two 511 keV annihilation photons together with a 1157 keV prompt gamma ($\gamma_p$). The primary interaction of incident photons in plastic scintillators occurs through Compton scattering depositing ranges of energies based on the scattering angle~\cite{NIM2014}. The TOT measurement in each scintillator strip is used to determine the incident photon energy deposition~\cite{Sharma:2020}. For a registered interaction (hit), the obtained TOT value results from averaging the TOT values measured from the matrix of Silicon Photomultipliers (SiPMs) attached at both ends of the scintillator and directly relates to the energy deposited by Compton-scattered photons. The measured TOT spectrum is shown in Figure \ref{selection}(A). The Compton edges  for 511 keV and 1157 keV photons appear at 7.5~$\mathrm{ns\cdot V}$ and 11~$\mathrm{ns\cdot V}$, respectively. These values correspond to the maximum energy transfer during a single Compton scattering event. The TOT signal range from 5.5 to 8 $\mathrm{ns\cdot V}$ is used to select 511 keV photons (red shaded area) and the TOT signal range from  8.1 to 14 $\mathrm{ns\cdot V}$ identifies 1157 keV photons (blue shaded area).  The formation of individual events relies on photon interactions (hits) that fall within a 20 ns coincidence  window for both $\gamma_a$ and $\gamma_p$. The selection criteria for 2-hit (2$\gamma_a$) events, which require exactly two 511 keV hits,  are explained in~\cite{das2024}. This work focuses on 3-hit (2$\gamma_a$+$\gamma_p$) events from the perspective of PLI. However, the comparison between reconstructed images based on events with a multiplicity of 2 hits (2$\gamma_a$) and 3-hits (2$\gamma_a$+$\gamma_p$) are presented.

The next selection criterion is based on the event-wise hit multiplicity (Fig.~\ref{selection}(B)). For PLI with $^{44}$Sc, we require exactly three hits per event ($\mu$=3): two corresponding to the 511 keV annihilation photons with $\mathrm{TOT}_{\mathrm{hit}}\in[5.5-8] \mathrm{ns\cdot V}$ and one corresponding to the 1157 keV prompt gamma with $\mathrm{TOT}_{\mathrm{hit}}\in[8.1-14]\ \mathrm{ns\cdot V}$. Fig.~\ref{selection}(B) illustrates the distribution of hit multiplicity ($\mu$) with the red-shaded region highlighting the events with exactly two annihilation photons and one prompt gamma, identified based on their $\mathrm{TOT}_{\mathrm{hit}}$ values. For events with a multiplicity of $\mu = 3$, the three hits are required to consist of two in the annihilation region and one in the prompt region of the $\mathrm{TOT}_{\mathrm{hit}}$ distribution. For events with $\mu > 3$, only those with two hits in the annihilation region and one in the prompt region are retained, while any additional hits with $\mathrm{TOT}_{\mathrm{hit}} < 5.5 \, \mathrm{ns\cdot V}$ or $\mathrm{TOT}_{\mathrm{hit}} \in (8-8.1) \, \mathrm{ns\cdot V}$ are excluded from the analysis.

To suppress the background from scattered photons or accidental coincidences, we imposed angular constraints on the emitting direction of photons. Fig.\ref{selection}(C) shows the distribution of the angle $\theta_{AA}$ between the annihilation photons calculated from their hit positions with respect to the detector center. Only events with $\theta_{AA}$ (red-shaded region in Fig.\ref{selection}(C)) were accepted for further analysis. Moreover, the prompt gamma $\gamma_d$ is emitted isotropically with respect to the annihilation photons. However the observed distribution of the angle between the annihilation photons and prompt gamma is strongly picked for small angles. This is due to the misidentification of  secondary scattering of photons in the detector as signals from prompt gamma or annihilation photon. Therefore, the selected events were further reduced by implementing the additional constraint ($\theta_{DA} \geq 30^\circ$) that significantly improved the purity of true event (Fig.\ref{selection}(D)).

Finally, we applied a Scatter Test (ST) to distinguish events with true annihilation photons from contamination by intra-detector Compton scatter or accidental coincidences. For annihilation photons registered with hit times $t_2$ and $t_3$, and hit positions $\vec{r}_2$ and $\vec{r}_3$, the ST is defined as 

\begin{equation}
\text{ST} = |t_3 - t_2| - |\vec{r}_3 - \vec{r}_2|/c,
\end{equation}
where $c$ is the speed of light.

In the ideal case of direct detection of annihilation photons ($t_2=t_3$), the value of ST is smaller than zero. In the case if the registered hits correspond to the primary and secondary scattering of the same photon, then ST = 0 within the experimental resolution. For the accidental coincidences, the time difference between the hits $|t_3 - t_2|$ may be large, yielding the ST value greater than zero. By selecting the annihilation pairs for which the ST value is less than (ST $<-0.5$~ns) the purity of the selected events was enhanced.
\subsection{Image reconstruction}\label{subsec3}
The reconstruction of images was performed for both conventional PET (2$\gamma_a$) and positronium (2$\gamma_a$+$\gamma_p$) events, enabling a direct comparison between standard and Ps-enhanced modalities. It is worth mentioning that in both cases only annihilation pairs (511~keV) were used for image reconstruction. In the category of 2$\gamma_a$+$\gamma_p$ events, an additional prompt gamma ($\gamma_p$) was included for subsequent lifetime analysis. 
The annihilation point $\vec{r}_a$ and annihilation time $t_{a}$ are calculated from the time and position of the annihilation hits (($t_2,\vec{r_2}$) and ($t_3,\vec{r_3}$)):

\begin{equation}\label{eq:posRecoFormula}
\vec{r}_a = \frac{\vec{r}_2 + \vec{r}_3}{2} 
        + \frac{c\left(t_3 - t_2\right)}{2} 
          \cdot \frac{\vec{r}_2 - \vec{r}_3}{|\vec{r}_2 - \vec{r}_3|},
\end{equation}
and
\begin{equation}
t_{a} = \frac{t_2 + t_3}{2} - \frac{|\vec{r}_2 - \vec{r}_3|}{2c}.
\end{equation}

As the next step, the data acquired with J-PET were converted into a list-mode file using the J-PET framework for CASToR, an open-source reconstruction software~\cite{merlin2018castor}. The image reconstruction used the Maximum Likelihood Expectation Maximization (MLEM) algorithm, which ran for 10 iterations. The dimensions of the reconstructed image were established at 200 × 200 × 200 voxels, with each voxel having a size of 2.5 × 2.5 × 2.5 mm$^3$. The reconstructed images were smoothed using a 5-mm Gaussian filter uniformly. The reconstructed images of $2\gamma_a$ events and $2\gamma_a + \gamma_p$ events, obtained using the MLEM algorithm, along with the annihilation position distribution ($\vec{r}_a$) [Eq.\ref{eq:posRecoFormula}] for the $2\gamma_a + \gamma_p$ events, are presented in the results section.
\subsection{Positronium Lifetime estimation}\label{subsecp}

The emitted positrons in $\beta^+$ decay can annihilate either by direct interaction with a surrounding  electron ($e^-$) $-$ a process lasting on average about 388~ps~\cite{doi:10.1126/sciadv.abh4394} or by the formation of an $e^+e^-$~bound state known as Ps. Ps can be produced in two distinct ground states: para-Ps (pPs), which has a mean  lifetime of approximately 125 ps, and ortho-Ps (oPs), a longer-lived state with a mean lifetime in water of of about 1.83 ns~\cite{waterortho1}. Consequently, the measured positron lifetime can vary depending on the annihilation mechanism.

The positron lifetime, defined as $\Delta T = t_a - t_p$, represents the time difference between positron emission and its subsequent annihilation. Here, $t_a$ denotes the positron annihilation time, while $t_p$ approximates the positron emission time. In the $^{44}$Sc decay process, the prompt gamma is emitted on average 2.6~ps after the emission of the positron, according to its decay scheme presented in Fig.~\ref{introduction}(A). Therefore, the prompt gamma emission time ($t_p$) is used as an approximation for the positron emission time. The emission time is determined by subtracting the time of flight of prompt gamma from the time of its registration ($t_1$): 
\begin{equation}
t_p = t_1 - \frac{|\vec{r}_1 - \vec{r}_a|}{c},
\end{equation}
where $\vec{r}_1$ is the registered hit position of prompt gamma and $\vec{r}_a$ corresponds to the annihilation point reconstructed using Eq.~\ref{eq:posRecoFormula}. The short range of positrons from $^{44}$Sc supports our assumption that the prompt gamma originates from the annihilation site. The average distance between emission and annihilation is approximately 1.7~mm in tissues~\cite{ManishBAMS}.

\begin{figure*}[htbp]
    \centering
        \includegraphics*[width=1\textwidth]{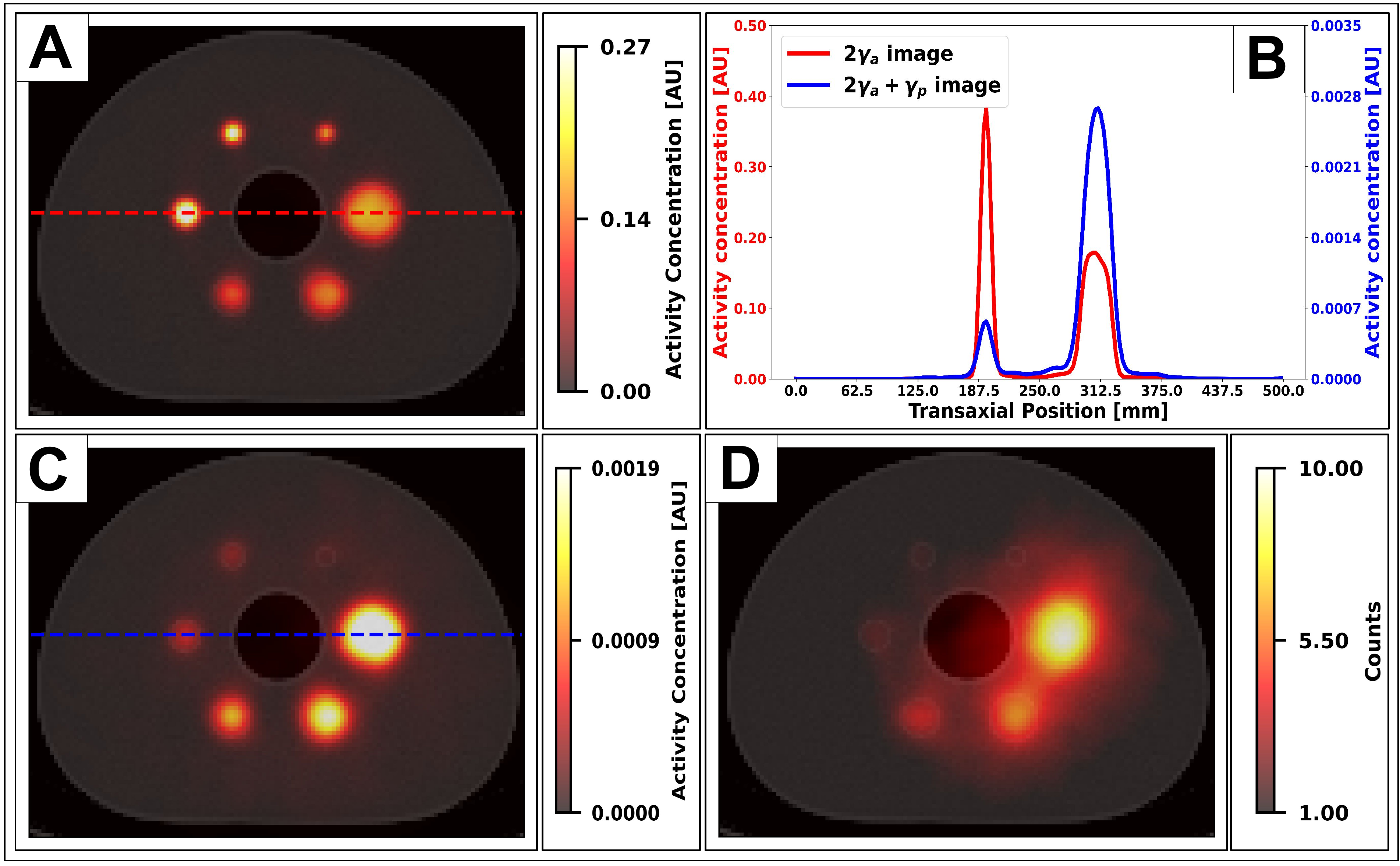}
    \caption{ {\bf{(A)}} Transaxial view of the conventional PET image ($2\gamma_a$) obtained from the modular J-PET, reconstructed with CASToR and overlaid on the CT image. {\bf{(B)}} Line profile along the indicated dashed line in the images. The red line represents the profile for the conventional PET image, showing that the $^{18}$F activity concentration is more than two times higher than $^{44}$Sc along the image slice. The blue line corresponds to the $2\gamma_a + \gamma_p$ image. {\bf{(C)}} Transaxial view of the $2\gamma_a$ image for $2\gamma_a + \gamma_p$ events, reconstructed using two 511 keV photons in CASToR and overlaid on the CT image. {\bf{(D)}} Transaxial view of the annihilation point distribution ($\vec{r}_a$) for $2\gamma_a + \gamma_p$ events overlaid on the CT image of the NEMA IQ phantom, with voxels having a relative intensity greater than 10\% displayed.
\label{image}
    }
\end{figure*}

The measured $\Delta T$ spectrum was analyzed by fitting a sum of exponential decay components convoluted with the experimental time resolution, using the dedicated PALS Avalanche software to estimate the mean oPs lifetime, as detailed in \cite{doi:10.1126/sciadv.abh4394, PALSAva1, PALSAva2, PALSAva3}. The initial values for pPs and direct-annihilation contributions were set to a lifetime of 125~ps with 10$\%$ relative intensity and 388~ps with 60$\%$ intensity, respectively. To optimize the fitting results, both lifetimes and their intensities were permitted to vary up to 200$\%$ of the initial value. The background level was estimated by averaging the bin counts in the range of -10~ns to -5~ns and subtracted before the fitting. Three distinct fitting approaches were used:

\noindent \textbf{Model 1 (Fixed Background):} maintained a constant background level at its pre-estimated value while allowing the oPs lifetime and intensity to vary freely.

\noindent \textbf{Model 2 (Constrained Background):} the background level was permitted to vary between -3$\sigma$ to +3$\sigma$ of its mean value where $\sigma$ represents the standard deviation of the counts in the -10~ns to -5~ns interval. The oPs lifeitme and intensity remained free. This method reduces bias and instability in low-statistics datasets through statistical fluctuation accommodation. 

\noindent \textbf{Model 3 (Extended Parameter Range):} the background remained constant while the oPs lifetime and intensity were permitted to change within 200$\%$ of their initial values which were set to 2~ns and 30$\%$, respectively.

In addition to the mean oPs lifetimes, the mean positron lifetime ($\Delta T_{\text{mean}}$) in the signal region from 0 to 5~ns in the $\Delta T$ spectrum was estimated after subtracting the background:
\begin{equation}
\Delta T_{\text{mean}}
  = \sum_{i=0~\mathrm{ns}}^{5\,\mathrm{ns}} (N_i-N_b)\Delta T_i\
    \bigl/\sum_{i=0~\mathrm{ns}}^{5\,\mathrm{ns}} {(N_i- N_b)},    
\end{equation}
where $N_i$ is the number of events for $\Delta T_i$ and $N_b$ is the number of background events estimated by averaging the bin counts in the range of -10~ns to -5~ns. The results of the estimated mean oPs lifetime ($\tau_\text{oPs}$) and $\Delta T_{mean}$ are presented in the Results~\ref{sec3} section.

\section{Results}\label{sec3}

The transaxial view of the MLEM reconstructed conventional PET ($2\gamma_a$) image, obtained with the J-PET scanner using the parameters described in the Methods section, is shown in Fig.~\ref{image}(A). The image is corrected for  sensitivity, attenuation, and normalization (see~\cite{das2024} for details). 
The effective activity concentration ratio in the reconstructed image, initially 3.10 for $^{18}$F to $^{44}$Sc, should be approximately 2.38 after a 178~minute acquisition due to the different isotopes' decay rates. This ratio reflects the time-integrated activities, calculated using decay constants of $0.006316~ \text{min}^{-1}$ for $^{18}$F and $0.002909~ \text{min}^{-1}$ for $^{44}$Sc over 178 minutes. The line profile shown in Fig.~\ref{image}(B) for the $2\gamma_a$ image indicates that the ratio of the maximum activity concentrations between $^{18}$F and $^{44}$Sc in the given image slice is approximately 2.08.

Next, the reconstructed $2\gamma_a + \gamma_p$ image using CASToR is shown in Fig.\ref{image}(C), demonstrating the effectiveness of the selection criteria. The line profile along the 17~mm and 37~mm diameter spheres [Fig.\ref{image}(B)] shows that events from $^{18}\text{F}$ are significantly reduced, as $^{18}\text{F}$ releases almost no prompt gammas associated with positron emission. The remaining $^{18}\text{F}$ events could be attributed to the possibility of detecting a prompt gamma from $^{44}\text{Sc}$ while registering two annihilation photons from $^{18}\text{F}$ within the 20~ns event time window, which introduces false $2\gamma_a + \gamma_p$ events in $^{18}\text{F}$, primarily contributing to the background.

\begin{figure}[htbp]
    \centering
        \includegraphics*[width=0.5\textwidth]{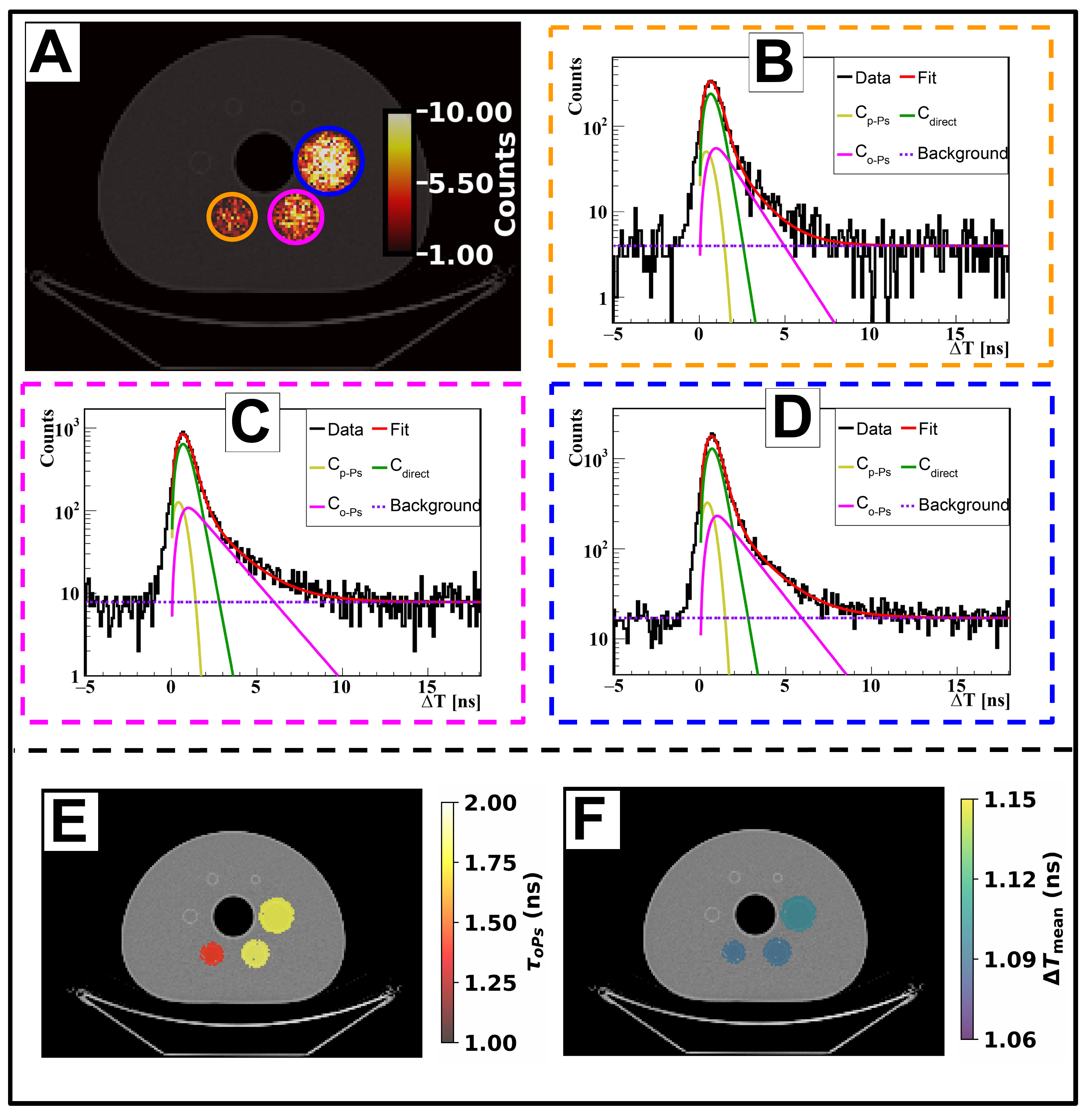}
    \caption{ \textbf{(A)} Transaxial view of the NEMA-IQ phantom with selected ROIs from the spheres, overlaid on the CT image of the NEMA IQ phantom. \textbf{(B–D)} Distributions of positron annihilation lifetimes ($\Delta T$) for the 22~mm (B), 28~mm (C), and 37~mm (D) diameter spheres. The black histograms represent the experimental data, while the overlaid curves correspond to the fitted components: pPs (C$_\text{pPs}$), direct annihilations (C$_\text{direct}$), oPs (C$_\text{oPs}$), and background from accidental coincidences. The red curve represents the total fit, obtained as the sum of all contributions. \textbf{(E–F)} Visualization of the estimated mean oPs lifetime ($\tau_\text{oPs}$) (E) and mean positron lifetime ($\Delta T_{\mathrm{mean}}$) (F) from Fitting Model 1, where the activity counts within each selected ROI are replaced by their respective lifetime values.
     \label{Result}
    }
\end{figure}

Furthermore, the annihilation position ($\vec{r}_a$) for $2\gamma_a + \gamma_p$ events, obtained using two 511 keV photons (Eq.~\ref{eq:posRecoFormula}), is presented in Figure~\ref{image}(D). The image clearly shows high counts in the spheres measuring 22~mm, 28~mm, and 37~mm in diameter, which were filled with $^{44}$Sc. The positron lifetime ($\Delta T$) values for each sphere were extracted by defining regions of interest (ROIs) within the different spheres, as shown in Fig. \ref{Result}(A), and were plotted using a bin width of 100 ps.

The body and spheres in the NEMA-IQ phantom are made of PMMA (polymethyl methacrylate), and the oPs lifetime in PMMA ranges from approximately 1.7 ns to 1.9 ns, depending on the temperature (20$^\circ$C to 60$^\circ$C) \cite{PMMAortho}. This range is comparable to the oPs lifetime in water, which is approximately 1.839 $\pm$~0.015~ns at 20$^\circ$C \cite{waterortho1}. Therefore, the contributions of oPs lifetime in water and PMMA were combined as a single component in the fitting procedure.

The results from the fit are presented in Table~\ref{table:fit} for the different fitting models. In Fitting Model 1, the mean oPs ($\tau_\text{oPs}$) lifetime in the 28 mm, 37 mm, and 22 mm diameter spheres is shown in Fig.~\ref{Result}(B-E). The mean oPs lifetime in the 28 mm and 37 mm spheres is in excellent agreement with the mean oPs lifetime in water ($1.839 \pm 0.015$~ns \cite{waterortho1}), while the mean oPs lifetime in the 22 mm sphere is $1.413 \pm 0.070$ ns. This discrepancy could be attributed to several factors: the number of events from this sphere was lower compared to the others, and the background might have been overestimated. To address this, in Fitting Model 2, the background level was allowed to vary within $\pm3\sigma$ of its estimated mean, resulting in a mean lifetime in the 22~mm diameter sphere that aligned more closely with the oPs lifetime in water. However, the mean oPs lifetime in the 28~mm sphere increased, while the lifetime in the 37~mm sphere remained consistent with the reference lifetime for water. Further validation of this model is needed, particularly to address the general challenges posed by low statistics conditions.

In Fitting Model 3, where all parameters were free within selected limits, the lifetime values remained consistent with those obtained from Fitting Model 1. 
Beyond these observations, the mean positron lifetime ($\Delta T_{\text{mean}}$) proved to be a more robust parameter for lifetime characterization, as previously demonstrated~\cite{doi:10.1126/sciadv.adp2840}. Notably, $\Delta T_{\text{mean}}$ is independent of the fitting model and depends only on the background estimation method used within each model. It showed consistent results within 10 ps across all background estimation methods, irrespective of variations in the mean oPs lifetime (as shown in Fig.~\ref{Result}(F) when the background is constant in Fitting Model 1 and Fitting Model 3).
\begin{table*}
\caption{Positronium lifetime and intensity obtained from fitting the positron lifetime spectra using three different models. $D$~represents the diameter of the spheres. $\tau_\text{oPs}$ is the mean lifetime of oPs. $I_\text{oPs}$, $I_\text{direct}$, and $I_\text{pPs}$ are the relative intensities corresponding to oPs annihilation, direct positron-electron annihilation, and pPs annihilation, respectively. $\Delta T_\mathrm{mean}$ denotes the mean positron lifetime.}
\centering
\small
\begin{tabular*}{\textwidth}{@{\extracolsep{\fill}} c c c c c c @{\extracolsep{\fill}}}
\toprule
\multicolumn{6}{c}{Fitting Model 1} \\
\midrule
{$D$ [mm]} & {$\tau_\text{oPs}$ [ns]} & {$I_\text{oPs}$ [\%]} & {$I_\text{direct}$ [\%]} & {$I_\text{pPs}$ [\%]} & {$\Delta T_\text{mean}$ [ns]} \\
\midrule
22.00 & 1.413 $\pm$ 0.070 & 28.46 $\pm$ 1.26 & 61.11 $\pm$ 1.43 & 10.43 $\pm$ 1.20 & 1.099 $\pm$ 0.013 \\
28.00 & 1.821 $\pm$ 0.061 & 25.55 $\pm$ 0.74 & 63.41 $\pm$ 0.89 & 11.04 $\pm$ 0.73 & 1.099 $\pm$ 0.008 \\
37.00 & 1.804 $\pm$ 0.042 & 25.84 $\pm$ 0.52 & 62.05 $\pm$ 0.63 & 12.10 $\pm$ 0.52 & 1.102 $\pm$ 0.005 \\
\midrule
\multicolumn{6}{c}{Fitting Model 2} \\
\midrule
{$D$ [mm]} & {$\tau_\text{oPs}$ [ns]} & {$I_\text{oPs}$ [\%]} & {$I_\text{direct}$ [\%]} & {$I_\text{pPs}$ [\%]} & {$\Delta T_\text{mean}$ [ns]} \\
\midrule
22.00 & 1.693 $\pm$ 0.087 & 25.68 $\pm$ 1.13 & 56.96 $\pm$ 1.38 & 17.36 $\pm$ 1.16 & 1.122 $\pm$ 0.013 \\
28.00 & 2.129 $\pm$ 0.076 & 23.43 $\pm$ 0.70 & 63.99 $\pm$ 0.88 & 12.58 $\pm$ 0.71 & 1.108 $\pm$ 0.008 \\
37.00 & 1.836 $\pm$ 0.044 & 25.40 $\pm$ 0.51 & 62.62 $\pm$ 0.62 & 11.98 $\pm$ 0.51 & 1.102 $\pm$ 0.005 \\
\midrule
\multicolumn{6}{c}{Fitting Model 3} \\
\midrule
{$D$ [mm]} & {$\tau_\text{oPs}$ [ns]} & {$I_\text{oPs}$ [\%]} & {$I_\text{direct}$ [\%]} & {$I_\text{pPs}$ [\%]} & {$\Delta T_\text{mean}$ [ns]} \\
\midrule
22.00 & 1.398 $\pm$ 0.068 & 29.13 $\pm$ 1.23 & 62.85 $\pm$ 1.38 & 8.02 $\pm$ 1.04 & 1.099 $\pm$ 0.013 \\
28.00 & 1.815 $\pm$ 0.008 & 25.68 $\pm$ 0.73 & 65.15 $\pm$ 0.89 & 9.17 $\pm$ 0.70 & 1.099 $\pm$ 0.008 \\
37.00 & 1.838 $\pm$ 0.044 & 25.23 $\pm$ 0.50 & 65.07 $\pm$ 0.60 & 9.71 $\pm$ 0.43 & 1.102 $\pm$ 0.005 \\
\bottomrule
\end{tabular*}
\label{table:fit}
\end{table*}

\section{Discussion}\label{sec4}

In this study, we performed the first PLI with the $^{44}$Sc using the modular J-PET scanner. The results show that $^{44}$Sc is a suitable radionuclide for PLI because it emits prompt gamma (1157~keV) after the emision of positron in almost 100\% cases, and has a half-life of 4.04 hours~\cite{Duran2022Half}. We measured oPs lifetimes in the spheres of the NEMA IQ phantom, where the mean oPs lifetime value in the two largest spheres of the phantom matches with the reported value in water \cite{waterortho1}. However, the results from the smallest sphere measurements highlight the statistical limitations and inaccuracies in the measurement of background estimation. In the analysis, we used different fitting models to mitigate these problems, as the wrong background estimation strongly affects the stability and accuracy of the mean oPs lifetime measurement.

We observed that, the application of background estimation constraints in Model 2 produced better consistency of the mean oPs lifetime ($\tau_\text{oPs}$) for the low-statistics data in the 22 mm sphere, thus showing the potential of adaptive background handling in such conditions. However, for larger sphere (28 mm) with higher statistics, it predicted a larger $\tau_\text{oPs}$ value, indicating that the flexibility in background estimation can also introduce inconsistencies. Therefore, the obtained results suggest that Model 2 could serve as a useful tool to evaluate background sensitivity in datasets with limited statistics. But, its application requires further validation before firm conclusions can be drawn.

Furthermore, the mean positron lifetime ($\Delta T_\mathrm{mean}$) was found to be a more robust parameter, varying within a 10~ps range across the background estimation methods applied in the fitting models, which confirms its significance for PLI.

The Modular J-PET system, built with cost-effective plastic scintillators~\cite{KAPLON2023168186}, is a promising proof of concept for scalable, total-body PET and positronium imaging. It operates in triggerless mode, allowing simultaneous multi-photon detection, which is essential for accurate positronium lifetime measurements. New research demonstrates how $^{44}$Sc serves a dual purpose in theranostics through diagnostic imaging and targeted radiotherapy, which makes it an ideal choice for multiple applications \cite{Moskal2024vision, Gomes2024Comparison}. The combination of the suitability of  $^{44}$Sc and the detection capability of modular J-PET establishes a strong experimental basis that may support the clinical and preclinical application of $^{44}$Sc-based PLI.
Furthermore, the enhanced large axial field-of-view (LAFOV) of total body PET systems will improve the sensitivity several times for PLI~\cite{doi:10.1126/sciadv.adp2840, pet-clinics}. Therefore, the advancement of PLI will lead to a broader medical application, for instance, exploring its potential as a novel diagnostic indicator \cite{doi:10.1126/sciadv.adp2840}.

\section{Conclusion}\label{sec5}

In this study, we successfully demonstrated the feasibility of PLI using $^{44}$Sc with the Modular J-PET tomograph carried out with a NEMA IQ phantom. The promising results of this study mark an important step forward and open new possibilities for advancing PLI techniques. With the advent of total body PET systems, PLI will become accessible through the enhancement of LAFOV scanners~\cite{Steinberger2024, Prenosil-quadra-2022, Moskal-NEMA-PMB-2021, Spencer-uExplorer-NEMA-2021} and hence the overall sensitivity of the scanner. Furthermore, the possibility of using the more enhanced and optimized reconstruction algorithm \cite{Huang2025Fast, Huang2025Statistical, Qi-positronium-2023, Berens2024analytic, Qi-positronium-2022, Chen2023, Shopa-Kamil-Bams2023, Huang_mic, Huangjnumed.125.270130, Chen2024}, will pave the way towards the application of PLI in clinical and preclinical scans.

\section*{Acknowledgements}

We acknowledge support from the National Science Centre of Poland through grants MAESTRO no. 2021/42/A/ST2/00423 (P.M.), OPUS no. 2021/43/B/ST2/02150 (P.M.), OPUS24+LAP no. 2022/47/I/NZ7/03112 (E.S.), SONATA no. 2023/50/E/ST2/00574 (S.S.) and SONATA no. 2020/38/E/ST2/00112 (E.P. del R.), the Ministry of Science and Higher Education through grant no. IAL/SP/596235/2023 (P.M.), the SciMat and qLife Priority Research Areas budget under the program Excellence Initiative – Research University at Jagiellonian University (P.M. and E.S.), the Research Support Module as part of the Excellence Initiative – Research University program at Jagiellonian University (M.D.). We also acknowledge Polish high-performance computing infrastructure PLGrid (HPC Center: ACK Cyfronet AGH) for providing computer facilities and support within computational grant no. PLG/2024/017688.\\

The authors declare the following financial interests/personal relationships which may be considered as potential competing interests with the work reported in this paper: Paweł Moskal is an inventor on a patent related to this work. [Patent nos.: (Poland) PL 227658, (Europe) EP 3039453, and (United States) US 9,851,456], filed (Poland) 30 August 2013, (Europe) 29 August 2014, and (United States) 29 August 2014; published (Poland) 23 January 2018, (Europe) 29 April 2020, and (United States) 26 December 2017. Other authors declare that they have no known conflicts of interest in terms of competing financial interests or personal relationships that could have an influence or are relevant to the work reported in this paper.

\bibliographystyle{IEEEtran}
\bibliography{sn-bibliography.bib}

\end{document}